# Simulation and analysis of a high-k electron scale turbulence diagnostic for MAST-U

D. C. Speirs[1], J. Ruiz Ruiz[2], M. Giacomin[3,4], V. H. Hall-Chen[5], A. D. R. Phelps[1], R. Vann[4], P. G. Huggard[6], H. Wang[6], A. Field[7] and K. Ronald[1]

[1] Department of Physics, SUPA, University of Strathclyde, Glasgow, G4 0NG, U.K.
[2] Rudolf Peierls Centre for Theoretical Physics, University of Oxford, Oxford OX1 3NP, U.K.
[3] Dipartimento di Fisica "G. Galilei", Università degli Studi di Padova, Padova, Italy
[4] York Plasma Institute, Department of Physics, University of York, Heslington, York, YO10 5DD, U.K.
[5] Institute of High Performance Computing, Agency for Science, Technology, and Research (A*STAR), Singapore 138632, Singapore.
[6] Millimetre Wave Technology Group, RAL Space, STFC Rutherford Appleton Laboratory, Didcot OX11 0QX, U.K.
[7] United Kingdom Atomic Energy Authority, Culham, United Kingdom of Great Britain and Northern Ireland.

E-mail: david.c.speirs@strath.ac.uk

**Abstract**

Plasma turbulence on disparate spatial and temporal scales plays a key role in defining the level of confinement achievable in tokamaks, with the development of reduced numerical models for cross-scale turbulence effects informed by experimental measurements an essential step. MAST-U is a well-equipped facility having instruments to measure ion and electron scale turbulence at the plasma edge. However, measurement of core electron scale turbulence is challenging, especially in H mode. Using a novel synthetic diagnostic approach, we present simulated measurement specifications of a proposed highly optimised mm-wave based collective scattering instrument for measuring both normal and binormal electron scale turbulence in the core and edge of MAST-U. A powerful modelling framework has been developed that combines beam-tracing techniques with gyrokinetic simulations to predict the sensitivity and spectral range of measurement, with a quasi-numerical approach used to analyse the corresponding instrument selectivity functions. For the reconstructed MAST 022769 shot, a maximum measurable normalised bi-normal wavenumber of $k_\perp \rho_e \sim 0.6$ was predicted in the core and $k_\perp \rho_e \sim 0.79$ near the pedestal, with localisation lengths $L_{FWHM}$ ranging from ~0.4 m in the core at $k_\perp \rho_e \sim 0.1$ to ~0.08m at $k_\perp \rho_e > 0.45$. Synthetic diagnostic analysis for the 022769 shot using CGYRO gyrokinetic simulation spectra reveal that ETG turbulence wavenumbers of peak spectral intensity comfortably fall within the measurable / detectable range of the instrument from the core to the pedestal. The proposed diagnostic opens up opportunities to study new regimes of turbulence and confinement, particularly in association with upcoming non-inductive, microwave based current drive experiments on MAST-U and can provide insight into cross-scale turbulence effects, while having suitability to operate during burning plasma scenarios on future reactors such as STEP (Spherical Tokamak for Energy Production).

Keywords: plasma, turbulence, scattering, gyrokinetic simulation

# 1. Introduction

## 1.1 Motivation

Heat transport, due to turbulence on the scale of the electron Larmor radius, is important in defining the confinement and consequent fusion performance of spherical aspect ratio tokamak plasmas [1, 2, 3]. In a tokamak, nested flux surfaces are formed by toroidal and poloidal magnetic fields. These surfaces permit relatively free transport of particles and energy along the lines of flux $\psi$ but limit cross-field transport in the radial direction[2]. This facilitates the development of a hot dense plasma core, with an energy confinement time long enough for ions and electrons to reach thermal equilibrium and temperatures sufficient for a sustainable fusion reaction [4]. Minimising cross-field transport is key to achieving the required plasma conditions. Turbulence dominates cross-field transport losses [1, 5]. Turbulent eddies exist at scales extending from the electron Larmor radius to the ion Larmor radius and macroscopic structures approaching the machine size. Although significant progress has been made in understanding the mechanisms and drivers of turbulent transport in plasma [1, 6, 7, 8], complex feedback mechanisms coupling the interaction of these eddies across many orders of magnitude in space and time make predictive modelling extremely challenging [9, 10]. Detailed experimental data at both electron and ion scales is therefore required to formulate the reduced models necessary to develop schemes that minimise turbulent cross-field transport.

Spherical aspect-ratio tokamaks, such as MAST-U at the Culham Centre for Fusion Energy (CCFE), offer a potential path to compact fusion power [11]. They benefit from enhanced confinement compared to larger conventional aspect-ratio tokamaks. This enhancement is believed to be caused by more extreme toroidicity and larger E×B shearing rates, both of which can supress electrostatic drift wave instabilities and turbulence at both ion and electron scales. Predictions of transport from discrete ion and electron scale simulations, however, do not always match experimental observations [12]. Numerically challenging cross-scale simulations, resolving both ion and electron contributions, reveal that the electron scale can indirectly enhance ion scale turbulence by disrupting sub ion-scale flows [9, 10]. This can lead to an order of magnitude increase in cross-field transport at the ion scale, significantly reducing confinement and the viability of achieving fusion ignition and burn in a spherical tokamak. With recent significant investments announced by governments [13, 14, 15, 16] and private industry [11] to develop tokamak based fusion powerplants, the urgency to understand cross-scale turbulent interactions has never been greater, with experimental measurements of electron and ion scale turbulence in the core plasma essential for model development. On MAST-U, the Beam Emission Spectroscopy (BES) diagnostic can measure ion scale turbulence in the core plasma [17]. The UCLA [18] and SWIP[19] DBS (Doppler Back-Scattering) diagnostics on MAST-U can measure intermediate to electron-scale turbulence up to $k_\perp \rho_s \sim 9$ ($k_\perp \rho_e \sim 0.15$) where $\rho_s = \sqrt{m_i T_i/(qB)}$ is the ion (deuterium) sound radius and $\rho_e$ the electron gyroradius, but have limited core accessibility (constrained to lower density operational regimes) and a measurement location and spatial localisation that is dependent on the cut-off surface location and gradient profile. We have therefore developed a novel, mm-wavelength coherent scattering diagnostic for MAST-U that can measure high-k (large wavenumber) electron scale turbulence under all operating conditions of the experimental reactor.

## 1.2 Proposed high-k scattering instrument

Spatial anisotropy has been observed in electron temperature gradient (ETG) driven turbulence which helps motivate the orientation / alignment of measurement for the high-k scattering diagnostic [20, 21, 22]. ETG electron scale turbulence is expected to be most significant in the bi-normal direction, i.e. perpendicular to the magnetic field and in-plane with the flux surface. Scale ranges are expected to be of order $k_\perp \rho_e \sim 0.1 \to 0.4$ [23] in the confinement region of the core plasma ($0.5 < r/a < 1$) where $k_\perp$ is the wavenumber of the turbulence, $r$ is the tokamak minor radial coordinate and $a$ is the tokamak minor radius. Multiscale simulations predict that turbulence growth rates peak at $k_\perp \rho_s \approx 0.45$ and $k_\perp \rho_s \approx 18$ ($k_\perp \rho_e \approx 0.3$) for ion and electron scales respectively [9, 10]. Simultaneous measurement of turbulence on both ion and electron scales within the core plasma of MAST-U using the BES and proposed high-k scattering diagnostics is therefore essential to validate numerical predictions. These measurements will be of particular high value on MAST-U, where a dual frequency electron Bernstein wave (EBW) heating and current drive system is scheduled for installation in 2024 [24]. Understanding the impact that different EBW and neutral-beam current-drive configurations have on plasma flow and associated confinement on MAST-U will be critical, and the ability to monitor variations in electron and ion scale turbulent transport within the core plasma will be of key importance for theory and model development.

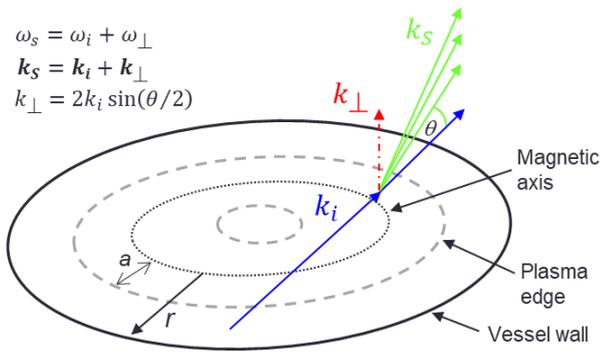

**Figure 1.** Proposed bi-normal high-k scattering geometry across MAST-U plasma. An equatorial plain representation shows the incident wavevector $\boldsymbol{k_i}$, turbulence wavevector $\boldsymbol{k_\perp}$ and scattered wavevector $\boldsymbol{k_s}$ along with three-wave matching and Bragg condition in terms of scattering angle θ.

The principle of a high-k scattering diagnostic is that part of a collimated electromagnetic wave is scattered at an angle given by the vector sum of its wavevector and that of the turbulence, with a Doppler shift in frequency due to the movement of the turbulent eddies: see Figure 1. Measurement of the deflected signal amplitude as a function of scattering angle allows the turbulence spectrum to be mapped in wavevector space, identifying the dominant scale lengths. Given an electron Larmor radius in MAST-U of ~0.2 mm, the implied turbulence wavelength range is of order a few mm. The scattering wavevector resolution scales as the reciprocal of the beam waist ($1/e^2$ intensity radius) at the measurement location, which is set to be of order ~2 cm. To measure unambiguously the scattered signals requires that they are spatially separated from the probe beam, and each other, by more than the beam width at the detector location. The probe beam frequency is set by a number of criteria: 1) it should be greater than twice the maximum prevailing plasma frequency to ensure unfettered access to the plasma volume with minimal effect of refraction on the beam path, 2) it should not be so high that the deflection associated with the turbulence wavevector range of interest would not present clearly separated beams, 3) adequately powerful sources and sensitive detectors are available and 4) it should not be at a harmonic of the frequencies of the MAST-U high power EBW heating gyrotrons, at 28 GHz and 34.8 GHz. This all points to a frequency in the range 330 to 400 GHz. The practical choice is 376 GHz, which takes advantage of the availability of 94 GHz transistor power amplifiers for radar applications when frequency multiplied by four, i.e. two successive stages of frequency multiplication can be used deliver a 376 GHz beam.

The proposed instrument will be novel in several ways. Other high-k diagnostics have been implemented, with great success, in particular a system developed by University of California, Davis (UCD) for NSTX [25, 26] and a new system being developed by UCD for NSTX-U [27]. The UCD system deployed on NSTX used a vacuum tube source and was configured for radial scattering. The proposed system for NSTX-U will focus on bi-normal scattering and will use a frequency > 600 GHz from a molecular vapour laser. To make our system highly deployable we propose to use solid state multiplication of high frequency, high power, solid state sources to a sub-mm frequency of 376 GHz, exploiting a rapidly developing field of enabling technologies. These sources are largely unaffected by the magnetic fields of a tokamak, resulting in closer proximity to the vacuum vessel and lower transmission loss. We have designed a predominantly reflective optical system, thereby avoiding dielectric loss and standing wave problems. Entrance and exit windows will be of low hydroxyl fused quartz with minimum absorption and a thickness tuned for Fabry-Perot resonant transmission. A particularly innovative feature is that the linear array of detectors will rotate to track the bi-normal direction for all plausible tokamak equilibria at a range of radial scattering locations from the axis to the pedestal. This will a) minimise the number of tokamak shots required to map the turbulence and b) more importantly, allow simultaneous measurements using all bi-normally aligned channels to reveal temporal correlations – this is anticipated to allow deeper scientific study of the coupling across the electron scale turbulence spectrum. A further capability of the proposed diagnostic is the ability to measure both bi-normal and normal (perpendicular to the magnetic field, perpendicular to the flux surface) oriented turbulence. This can be achieved by off-setting the rotational pitch of the carriage from strictly bi-normal. Each channel will then be in alignment with a combination of bi-normal and normal scattering contributions. The precise ratio and magnitude of these contributions can be computed for a given equilibrium via beam tracing. Results of simulations presented later in this paper demonstrate this.

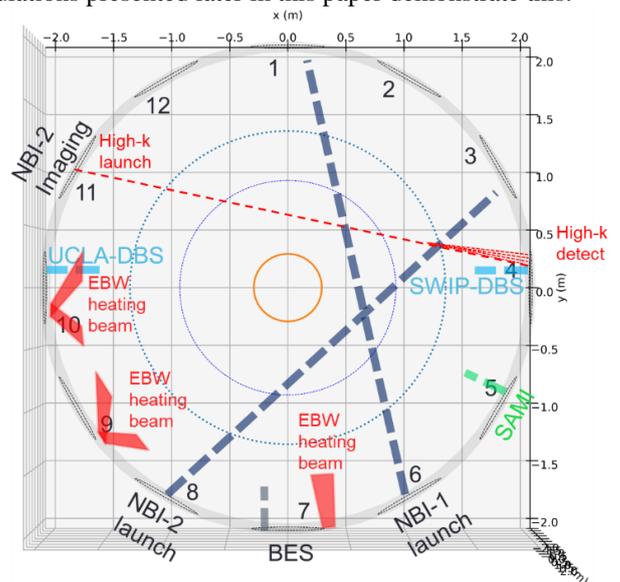

**Figure 2.** Proposed bi-normal high-k scattering geometry and installation across MAST-U vessel. Other key systems and diagnostics are illustrated for reference [17, 18, 19].

## 2. Theory

*2.1 Collective scattering*

The scattering of microwaves by density fluctuations in plasma has already proven a valuable method for diagnosing fusion plasma turbulence [26, 28]. Such scattering can be termed coherent (collective) if the turbulence wavelength is of the order of or much greater than the Debye length in the plasma $\lambda_{De} = \sqrt{(\epsilon_0 k_B T_e/n_e e^2)}$, where $k_B$ is Boltzmann constant, $\epsilon_0$ is the vacuum permittivity, $T_e$ is the electron temperature, $n_e$ is the electron density and $e$ is the electron charge. In a coherent scattering regime, the scattered waves are representative of the collective electron motion manifesting as fluctuations in the refractive index. The total scattered power for a plane wave scattered by a single coherent density fluctuation is given by the classical formula [25]

$$P_s = \frac{1}{4} P_i r_e^2 L_{FWHM}^2 \lambda_i^2 \delta n_e^2 \quad (1)$$

where $\lambda_i$ is the incident beam wavelength, $P_i$ is the incident beam power, $L_{FWHM}$ is the spatial resolution (full-width half-maximum overlap of scattered and incident beams enhanced by magnetic field pitch rotation with radius), $r_e$ is the classical electron radius and $\delta n_e$ is the density modulation amplitude. This formula makes no assumption about the anisotropy of a density fluctuation spectrum and provides an estimate of the minimum density fluctuation power that can be detected. In strongly magnetized plasmas, the spectrum of the turbulence is anisotropic in the parallel and perpendicular directions with respect to the background magnetic field. The turbulence exhibits length scales perpendicular to the magnetic field that are much smaller than along the field. The turbulence fluctuations are commonly described by a wavevector $k_\perp$ perpendicular to the field, while its component $k_\parallel$ is small, $k_\parallel \ll k_\perp$, consistent with the ordering in the turbulence theory of gyrokinetics [29, 30]. This means that in order to measure electron-scale fluctuations using high-k scattering, one needs to carefully design the launching and receiving apparatus in such a way that the incident and scattered rays not only intersect, but they do so while satisfying the condition that the difference between the scattered and incident beam wavenumbers at the scattering location, $\boldsymbol{k}_s - \boldsymbol{k}_i$, lies on the perpendicular plane to the magnetic field [31]. We call this difference the measured, or selected, turbulence wavenumber $\boldsymbol{k}_\perp = \boldsymbol{k}_s - \boldsymbol{k}_i$. In this manuscript, we have designed the incident and scattered beams such that this condition is satisfied for all the selected wavenumbers shown. We are particularly interested in the scattering of a microwave beam of radius $w_b$ from a turbulent, anisotropic density fluctuation spectrum $\delta \hat{n}_e(\boldsymbol{k}, \omega)$. A more appropriate formula for the scattered power over an angular aperture of $\pi(2/k_i w_b^2)$ for such a density fluctuation spectrum can be estimated by [32]

$$P_s = P_i \frac{k_i^2 L_{FWHM}^2}{k_\perp^2 w_b^2} \frac{\omega_{pe}^4}{\omega_i^4} \left(\frac{\delta n_e^2}{n_e^2}\right)_{rms} \quad (2)$$

where $\omega_{pe} = (n_e e^2/m_e \epsilon_0)^{\frac{1}{2}}$ is the electron plasma frequency, $\omega_i$ is the incident angular frequency, $k_i = \omega_i/c$ is the incident wavenumber of the microwave beam, $k_\perp = |\boldsymbol{k}_i - \boldsymbol{k}_s|$ is the magnitude of the selected scattered wavenumber of the turbulence, and $(\delta n_e/n_e)_{rms}^2$ is the root mean square of the turbulence fluctuation power. Equation [2] will be used in this manuscript to calculate the threshold $(\delta n_e/n_e)_{rms}^2$ that can be detectable by the diagnostic. Details of the derivation can be found in reference [29]. Note that in equation [2], the only unknown is the spatial resolution $L_{FWHM}$. An estimate of the diagnostic spatial resolution is given next.

Using an analysis following that of Mazzucato [33], one can obtain the instrument selectivity function for the receiving detector channel. We consider the beam spectrum $G(\kappa_\perp) = \exp(-\kappa_\perp^2/\Delta^2)$ where $\Delta = 2/w_b$ (assuming we are scattering from the beam waist) and $\kappa_\perp$ is the wavenumber perpendicular to the direction of propagation of the channel matched wavevector $\boldsymbol{k}_{S1}$, which satisfies the scattering condition that $\boldsymbol{k}_{S1} - \boldsymbol{k}_i$ remains perpendicular to the local magnetic field $\boldsymbol{B}$. This ensures that the scattered beam of wave vector $\boldsymbol{k}_{S1}$ arrives at the detector with maximal efficiency. In the proposed diagnostic, the detector and receiving optics have angular adjustment to ensure that binormal alignment is met for $\boldsymbol{k}_{S1}$ at the selected scattering coordinate. A beam of central wave vector $\boldsymbol{k}_{Sm}$ originating from the same scattering location however will result in $\boldsymbol{k}_{Sm} - \boldsymbol{k}_i$ having a component along $\boldsymbol{B}$. Therefore, $\boldsymbol{k}_{Sm}$ will be mismatched with respect to $\boldsymbol{k}_{S1}$, and the scattered amplitude arriving at the detector will be attenuated by the mismatch. This property has been extensively used in the past to localize high-k scattering measurements [31, 33] as well as more recently in DBS [34, 35, 36] while $\boldsymbol{k}_{Sm}$ is the mismatched wavevector (see Figure 3).

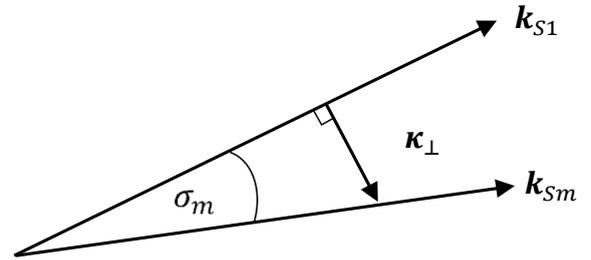

Figure 3: Matched $\boldsymbol{k}_{S1}$ and mismatched $\boldsymbol{k}_{Sm}$ scattered beam wavevectors with mismatch angle $\sigma_m$.

In what follows, we closely follow Mazzucato [33, 37] and use the property of the mismatched wave vectors to arrive at an expression for the instrument selectivity function $F$. We start by calculating the projection of the mismatched scattered wavevector in the perpendicular plane $\kappa_\perp$ of the matched wavevector

$$\boldsymbol{\kappa}_{\perp Sm} = \boldsymbol{k}_{Sm} - (\boldsymbol{k}_{Sm} \cdot \hat{\boldsymbol{s}}_1)\hat{\boldsymbol{s}}_1$$

After squaring both sides and some vector algebra we get

$$\boldsymbol{\kappa}_{\perp Sm}^2 = \boldsymbol{k}_{Sm}^2 (1 - (\hat{\boldsymbol{s}}_1 \cdot \hat{\boldsymbol{s}}_m)^2)$$

where $\hat{\boldsymbol{s}}_1$ and $\hat{\boldsymbol{s}}_m$ are the unit vectors for the matched and mismatched scattered beams respectively and $\hat{\boldsymbol{s}}_1 \cdot \hat{\boldsymbol{s}}_m = \cos \sigma_m = \frac{\boldsymbol{k}_{S1} \cdot \boldsymbol{k}_{Sm}}{|\boldsymbol{k}_{S1}||\boldsymbol{k}_{Sm}|}$. Substituting into the equation for the beam spectrum $G(\boldsymbol{\kappa}_\perp)$ we have

$$F = G(\boldsymbol{\kappa}_{\perp Sm}) = \exp\left(\frac{-k_i^2}{\Delta^2}\left(1 - \left(\frac{\boldsymbol{k}_{S1} \cdot \boldsymbol{k}_{Sm}}{|\boldsymbol{k}_{S1}||\boldsymbol{k}_{Sm}|}\right)^2\right)\right) \quad (3)$$

where $F$ is the instrument selectivity function, which can be analysed via a quasi-numerical approach using the projected wavevectors of mismatched scattered beams. The unit vectors $\hat{\boldsymbol{s}}_1$ and $\hat{\boldsymbol{s}}_m$ The mismatch is due to magnetic field pitch rotation relative to the value at the scattering coordinate, resulting in rotational misalignment of the scattered beam with respected to matched incidence.

Another fundamental limit on the instrument localisation of measurement is the projection of the Gaussian scattered beam intensity relative to the primary beam – the so-called "beam overlap". The impact of this is illustrated in Figure 4 and has been previously analysed for a collective Thomson scattering diagnostic [38].

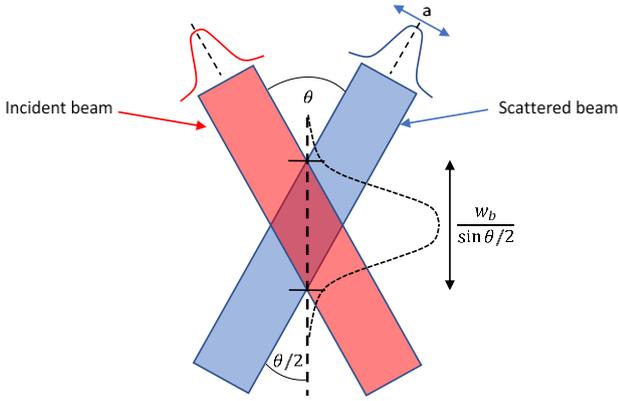

**Figure 4.** Illustration of incident and scattered Gaussian beam overlap along coordinate L, with the $1/e^2$ scattered beam waist projected at θ/2 to the incident beam, where θ is the scattering angle, yielding an effective $1/e^2$ overlap radius of $w_b$ / sin (θ/2).

With reference to Figure 4, the Gaussian beam waist overlap is maximum at an angle of $\theta/2$ to the primary and scattered beams, yielding an overlap waist of $w_b/\sin(\theta/2)$ and an instrument selectivity envelope of

$$F_{Gauss}(L) = \exp\left(-\left(\frac{L\Delta \sin(\theta/2)}{2}\right)^2\right) \quad (4)$$

Combined with the mismatch instrument selectivity function in (3), we have a resultant instrument selectivity function of

$$F(L) = \exp\left(\frac{-k_i^2}{\Delta^2}\left(1 - \left(\frac{\boldsymbol{k}_{S1} \cdot \boldsymbol{k}_{Sm}}{|\boldsymbol{k}_{S1}||\boldsymbol{k}_{Sm}|}\right)^2\right) - \left(\frac{L\Delta \sin(\theta/2)}{2}\right)^2\right) \quad (5)$$

The first term in the exponential of equation 5 governs the localisation effect due to variation in magnetic field pitch angle rotation with radius, whilst the second term accounts for the incident and scattered Gaussian beam overlap region as a function of scattering angle $\theta/2$. From equation 5 the spatial localisation length $L_{FWHM}$ can be estimated by taking the FWHM of the resultant peak profile centred around the scattering coordinate. This will provide a quantitative assessment of the minimum scattered power detectable by the diagnostic for a given $\delta n_{e_{rms}}^2$. The spatial resolution $L_{FWHM}$ will also be used in a synthetic high-k diagnostic to quantitatively predict the scattered power spectrum.

The selected $k_\perp$ is routinely decomposed into its components in the directions normal and bi-normal to the background magnetic field $B = B\,\hat{\mathbf{b}}$, where $\hat{\mathbf{b}}$ is the unit vector along the background magnetic field and $B$ is its magnitude. The normal direction to the flux surface is directed along the normal unit vector $\hat{\mathbf{e}}_n = \frac{\nabla\psi}{|\nabla\psi|}$, where $\nabla\psi$ is the gradient of the flux function. The bi-normal direction is directed along the bi-normal unit vector $\hat{\mathbf{e}}_b = \hat{\mathbf{e}}_n \times \hat{\mathbf{b}}$. Using these definitions, we define the normal and bi-normal wavenumber components of the selected wavevector $\boldsymbol{k}_\perp$ by $\boldsymbol{k}_\perp = k_n\hat{\mathbf{e}}_n + k_b\hat{\mathbf{e}}_b$. The normal and bi-normal wavenumber components of the turbulence will be used in section 3.4 to implement a synthetic diagnostic for high-k scattering.

## 2.2 Beam tracing of primary and scattered waves

The beam tracing code Scotty [36] has been used to compute the primary and scattered ray trajectories (receiving window aperture limited) as a function of radius for a variety of MAST-U operational equilibria. Scotty is a beam tracing code written entirely in Python 3 using cylindrical polar coordinates $(R, \zeta, Z,)$ natively. This simplifies the beam tracing equations by exploiting the toroidal symmetry of tokamaks. Scotty assumes lossless propagation and was executed without relativistic corrections to the electron mass, a mode suitable for lower temperature devices such as MAST-U.

Scotty solves the beam tracing equations for the primary ray trajectory, projecting from the launching port to the receiving port, through the plasma. Scattered ray trajectories are then computed for an array of prescribed bi-normal $k_b$ and normal $k_n$ turbulence wavenumbers. Looking at Figure 5, the turbulence wavevector projections are presented in cylindrical polar coordinates in $R-z$ and $\zeta-z$ planes along with the magnetic field vector **B**, where $\zeta$ is the toroidal coordinate, $R$ is the radial coordinate and $z$ is the vertical coordinate (perpendicular to the equatorial plane). The angles $\phi_{Rz}$ and $\phi_{\zeta z}$ are respectively the $R-z$ and $\zeta-z$ angles of the magnetic field at a given scattering coordinate in the plasma.

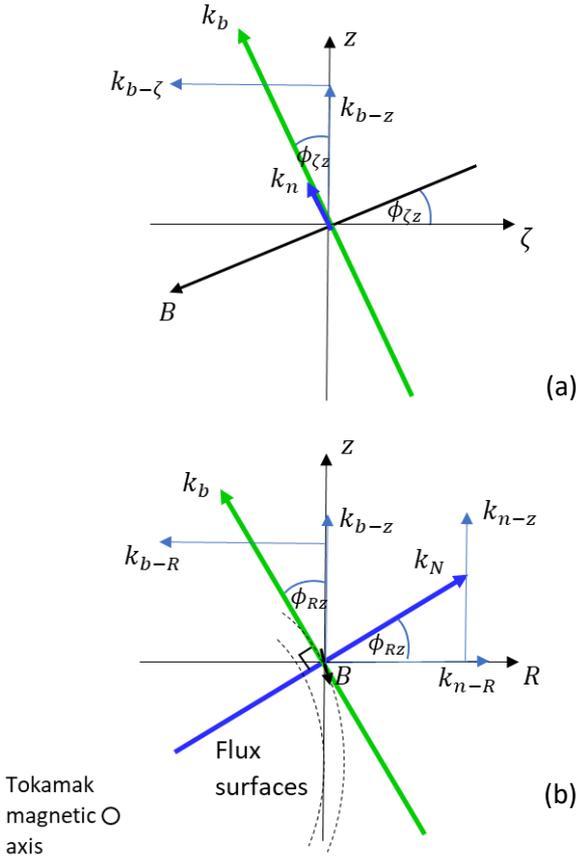

**Figure 5.** Scattering wavenumber projections in cylindrical polar coordinates for (a) R-z and (b) ζ-z planes.

Scotty solves the beam tracing equations for the primary ray trajectory, projecting from the launching port to the receiving port, through the plasma. Scattered ray trajectories are then computed for an array of prescribed bi-normal $k_b$ and normal $k_n$ turbulence wavenumbers. Looking at Figure 5, the turbulence wavevector projections are presented in cylindrical polar coordinates in $R-z$ and $\zeta-z$ planes along with the magnetic field vector **B**, where $\zeta$ is the toroidal coordinate, $R$ is the radial coordinate and $z$ is the vertical coordinate (perpendicular to the equatorial plane). The angles $\phi_{Rz}$ and

$\phi_{\zeta z}$ are respectively the $R-z$ and $\zeta-z$ angles of the magnetic field at a given scattering coordinate in the plasma. As the scattering wavenumber $\boldsymbol{k}_s = \boldsymbol{k}_i + \boldsymbol{k}_\perp$ where $\boldsymbol{k}_i$ is the incident beam wavenumber and $\boldsymbol{k}_\perp$ is the turbulence wavenumber, the contributions of $k_b$ and $k_n$ must be calculated in cylindrical polar coordinates, $k_R$, $k_\zeta$ and $k_z$, subject to the angular rotations $\phi_{Rz}$ and $\phi_{\zeta z}$. This yields the following set of scattering equations in $k_n$ and $k_b$

$$k_R = k_{R_p} + S_b \sin \phi_{Rz} k_b + S_n \cos \phi_{Rz} k_n$$
$$k_\zeta = k_{\zeta_p} + S_b \sin \phi_{\zeta z} \cos \phi_{Rz} k_b - S_n \sin \phi_{\zeta z} \sin \phi_{Rz} k_n \quad (6)$$
$$k_z = k_{z_p} + S_b \cos \phi_{\zeta z} \cos \phi_{Rz} k_b - S_n \cos \phi_{\zeta z} \sin \phi_{Rz} k_n$$

where $k_\zeta$ is negative for clockwise toroidal rotation and positive for counter-clockwise toroidal rotation in Scotty (viewed from above), $k_R$ is positive for radially outwards components and negative for radially inwards components and $k_z$ is positive for upwards (above midplane) components and negative for downwards (below midplane) components. The parameter $S_n = 1$ is the scattering sign (direction) for $k_n$ and $S_b = -1$ is the scattering sign (direction) for $k_b$.

Looking at Figure 6, one can see the resultant scattered component distribution for a predefined range in $k_b$ and $k_n$. The primary ray is illustrated in red, the green scattered components represent strictly bi-normal scattering from $k_b = 3.98 \times 10^2 \to 1.78 \times 10^3$ m$^{-1}$, whilst the blue scattered components correspond to a $k_b = 3.98 \times 10^2$ m$^{-1}$ and $k_n = 8.75 \times 10^2 \to 3.5 \times 10^3$ m$^{-1}$. The arrows indicate the respective directions of the magnetic field, bi-normal and normal wavevectors.

## 3. Results and analysis

### 3.1 Optical component configuration

The optical component configuration of the launching and receiving carriages for the proposed high-k diagnostic is now considered via ABCD matrix analysis [39]. The proposed high-k scattering instrument comprises a 376 GHz Gaussian beam launched across the MAST-U vacuum vessel between opposing equatorial ports, at near perpendicular incidence to the magnetic field and toroidal coordinate at the plasma outboard pedestal. The Gaussian beam divergence and $1/e^2$ radius is controlled to ensure a flat beam waist region of ~ 2 cm from the magnetic axis to the outboard plasma pedestal, closest to the receiving window. This facilitates adjustment of the scattering radius over the full depth of the plasma and normalised radial range of $r/a = 0 \to 1$, whilst maintaining a constant $1/e^2$ radius. A projection of the beam waist evolution is given in Figure 7 from an ABCD matrix code. One spherical focussing mirror with $R_{c1} = 0.635$ m is employed to focus the divergent beam following the launching antenna, with a planar 2-axis adjustable steering mirror inserted prior to traversing the vacuum vessel wall. At the receiving end, a focussing-defocussing mirror pair with radii of curvature $R_{c2} = 0.80$ m and $R_{c3} = -0.60$ m respectively is used to focus and redirect the beam into the detector array with a $1/e^2$ waist radius $w_b \approx 0.8$ cm.

An illustration of the launching optics carriage is presented in Figure 8. This will be mounted 0.16 m above midplane, horizontally aligned on a 600 mm equatorial port flange, utilising a 120 mm diameter low hydroxyl fused silica window for minimum transmission losses on entry into the vacuum vessel. A rotatable linear polariser will facilitate precise polarisation control of the launched beam. The detection optics carriage is illustrated in Figure 7. The receiving optics will collect the primary and scattered beams via a 250mm × 290mm elliptical window, also fabricated from low OH content fused silica. The thickness of this window will be tuned for maximum Fabry-Perot resonant transmission corresponding to the wavenumber with lowest scattered power (largest scattering angle), maximising the signal to noise ratio for the weakest signals. The entire assembly within the receiving carriage is mounted on a linear translation stage, allowing the focus of the receiving optics, i.e. the scattering volume, to be scanned in radius via a motorised linear drive. A separate rotational stage facilitates rotation of the receiving optics assembly and linearly aligned detector channel array. This rotation is centred around the primary ray, enabling alignment to be maintained with the poloidal scattering direction dictated by the pitch angle of the magnetic field, which varies as a function of radius and under different operational equilibria. A linear polariser is also mounted on entry into the receiving optics carriage. This can be independently rotated to facilitate co and cross-polar detection of the scattered radiation, allowing detection of both density and magnetic fluctuations within the scattering volume.

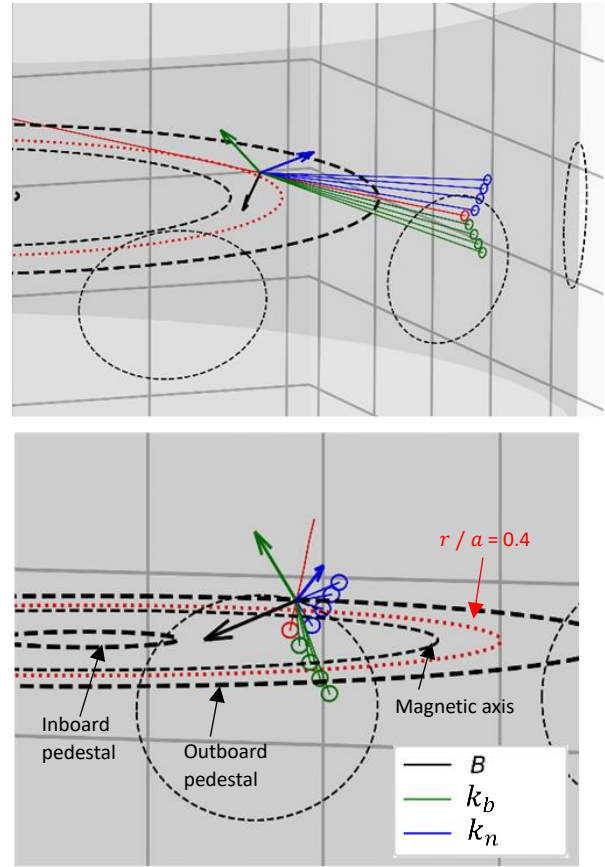

**Figure 6.** Scotty ray-tracing projections of bi-normal (green) and normal (blue) turbulence scattering from a reference beam (red). The scattered spectrum turbulence wavenumers are $k_n = 0$, $k_b = 3.98 \times 10^2$, $8.59 \times 10^2$, $1.32 \times 10^3$, $1.78 \times 10^3$ m$^{-1}$(green) and $k_b = 3.98 \times 10^2$ m$^{-1}$, $k_n = 8.75 \times 10^2$, $1.75 \times 10^3$, $2.63 \times 10^3$, $3.5 \times 10^3$ m$^{-1}$ (blue).

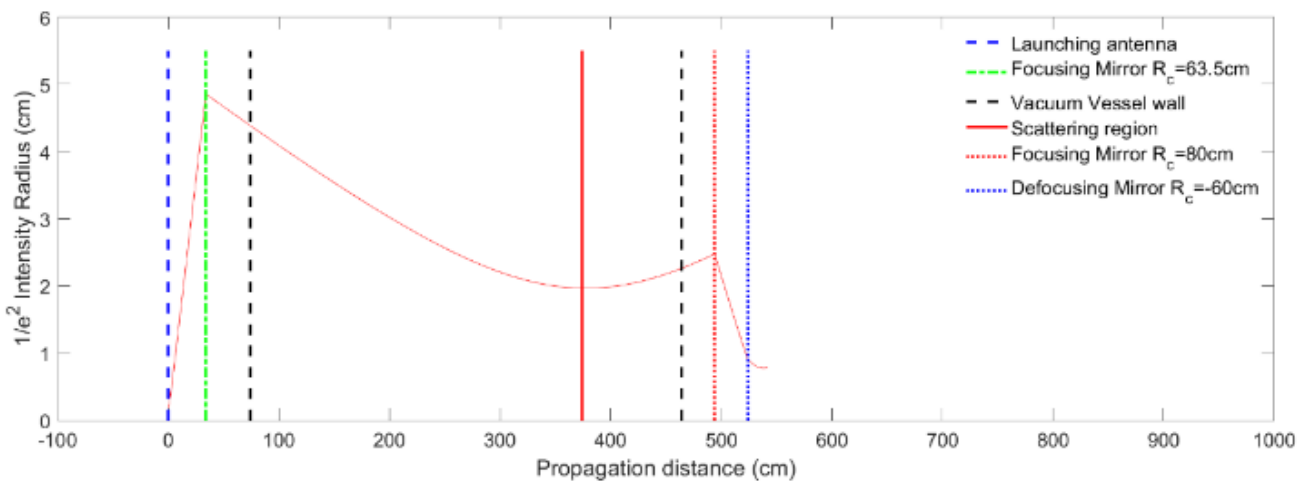

**Figure 7.** ABCD matrix calculations of Gaussian beam waist evolution for primary beam projected through launching and receiving optics and across the MAST-U vacuum vessel.

The blue dashed lines in the lower-right quadrant of Figure 9 indicate the positions of the upper and lower P5 poloidal field coils in MAST-U. The position of the elliptical receiving window and path of the scattered rays has been optimised to ensure no interception of scattered components on the P5 coils or their mounting brackets.

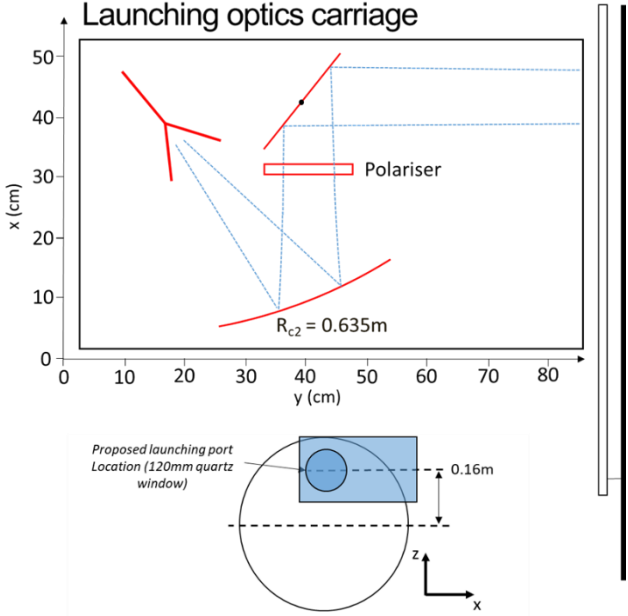

**Figure 8.** Illustration of launching optics carriage including launching antenna, focussing mirror, 2-axis adjustable redirecting mirror and rotatable linear polariser.

*3.2 Ray tracing simulations of the scattered spectrum and analysis of the instrument selectivity function*

The beam tracing code Scotty [36] has been used in ray tracing mode to predict the aperture limited primary and scattered beam paths for the high-k diagnostic as a function of scattering radius and operational equilibrium on MAST-U. The diametrically opposing ports on MAST-U have been selected for this analysis due to their optimal positioning to achieve a primary beam path projected across the plasma that is near-perpendicular to the magnetic field and outer pedestal over the scattering region, while avoiding tangential propagation near the inner pedestal earlier in the trajectory that would result in greater refraction of the primary beam path. Figure 10 shows the primary and scattered ray propagation paths for a sample high-beta MAST-U equilibrium (see appendix) with an aperture limited 4 channel scattered component distribution for three radial scattering coordinates. The ray trajectories are shown in Figure 10a, with the corresponding magnetic field pitch angle rotations in $\zeta - z$ and $r - z$ plotted as a function of radius and the scattering coordinate values highlighted.

The maximum $k_\perp \rho_e$ values measurable for each scattering radius (aperture limited) are shown, with a maximum $k_\perp \rho_e$ of 0.43 at $R_{scatt} = 1.24$ m. Figure 10b shows the localisation length of the scattering region as a function of $k_\perp \rho_e$ scattering radius. The localisation lengths $L_{FWHM}$ are the FWHM of the instrument selectivity functions plotted in Figure 10c which are generated using equation 5, showing the overlap of the scattered and incident Gaussian beam envelopes [38] constrained by the pitch rotation of the magnetic field [37] along the beam overlap length $L$ (at $\theta/2$ to the primary and scattered rays). The localisation length has an upper limit corresponding to the last closed flux surface at $R = 1.3$ m.

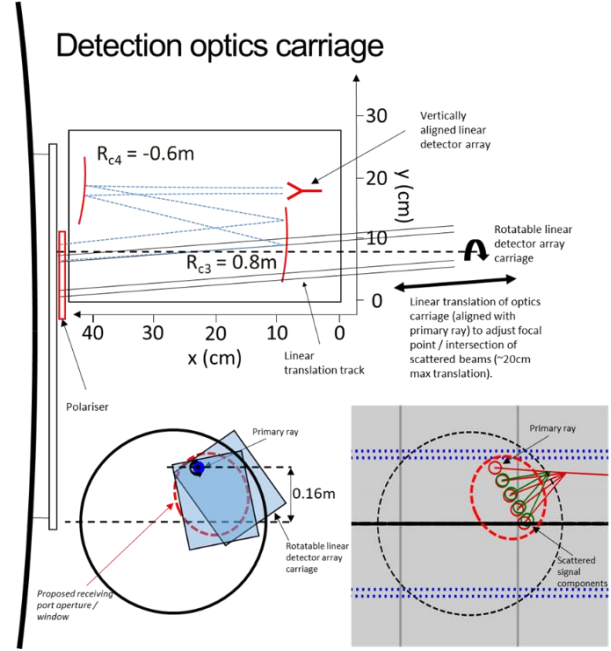

**Figure 9.** Illustration of receiving optics carriage including focussing / defocussing mirror pair, rotatable linear polariser and vertically aligned receiving antennae array.

Figure 11 shows the receiving window (aperture limited) 4 channel scattered component distributions for the 022769 MAST shot [41]. There are 4 radial scattering coordinates used in this case, corresponding to $r/a = 0.4, 0.5, 0.6$ and $0.8$. The scattering data from these simulations is used in the synthetic diagnostic analysis that follows in section 3.4, using the results of gyrokinetic simulations of ETG turbulence for the corresponding $r/a$ values outlined in section 3.3 below. Due to the lower magnetic field used in the 022769 shot when compared with the MAST-U high-beta sample equilibrium, the corresponding aperture limited $k_\perp \rho_e$ range is larger for equivalent scattering radii, with a maximum $k_\perp \rho_e$ of 0.79 for $r/a = 0.8$ ($R_{scatt} = 1.268$ m). Figure 12a and 12b shows the scattered signal power to noise ratio as a function of $k_\perp \rho_e$ for the MAST-U high-beta sample equilibrium and MAST 022769 shot respectively. These were calculated using the localisation data $L_{FWHM}$ plotted in Figures 10 and 11 substituted into equation 1 to compute the scattered power for a plane wave scattering off a coherent density fluctuation.

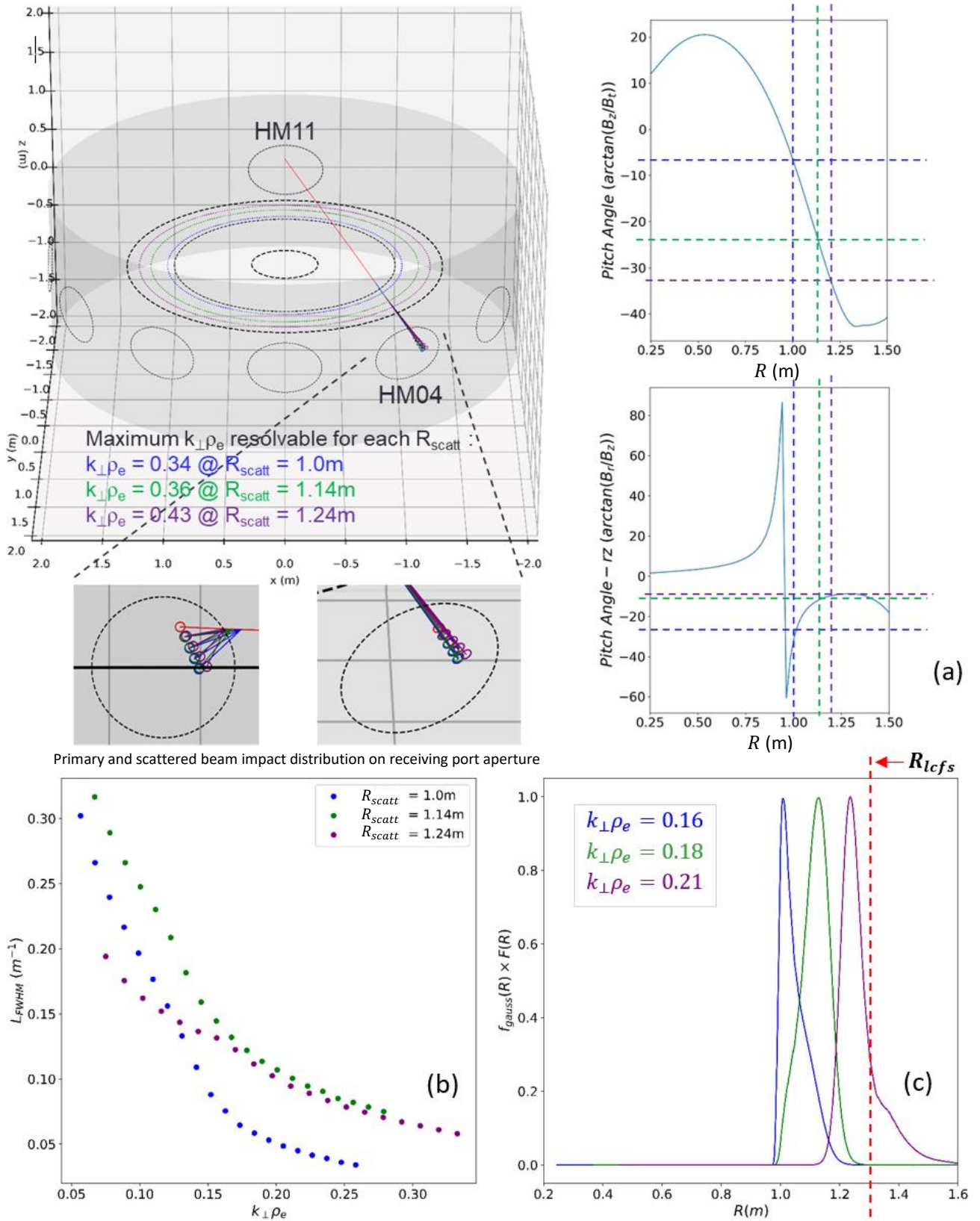

**Figure 10.** (a) High-k poloidal scattering projections for a simulated MAST-U high-beta sample equilibrium. (b) Localisation estimates as a function of $k_\perp \rho_e$ for three scattering radial coordinates corresponding to the FWHM of (c) the instrument selectivity functions.

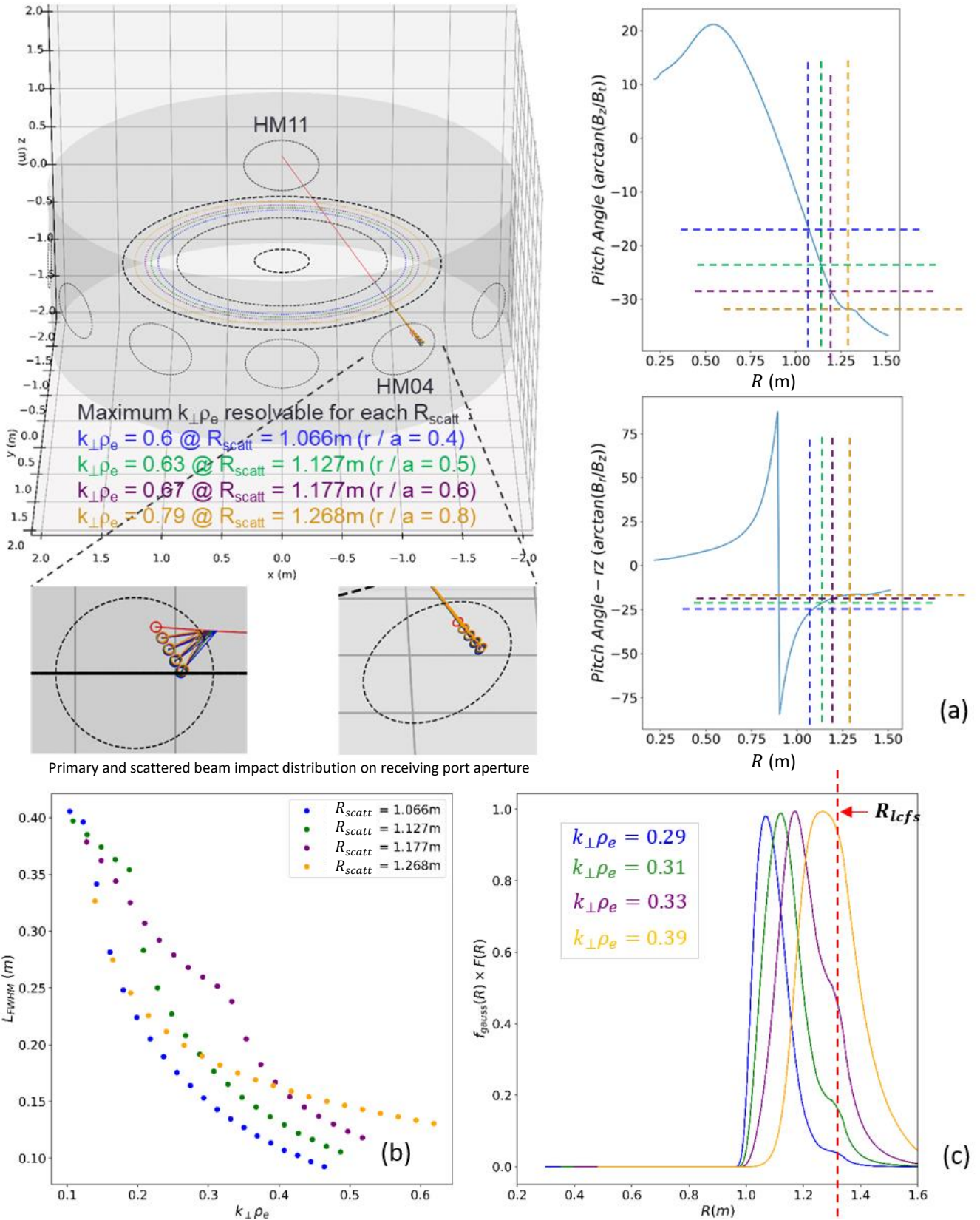

**Figure 11.** (a) High-k poloidal scattering projections for a simulated MAST-U high-beta sample equilibrium. (b) Localisation estimates as a function of $k_\perp \rho_e$ for three scattering radial coordinates corresponding to the FWHM of (c) the instrument selectivity functions.

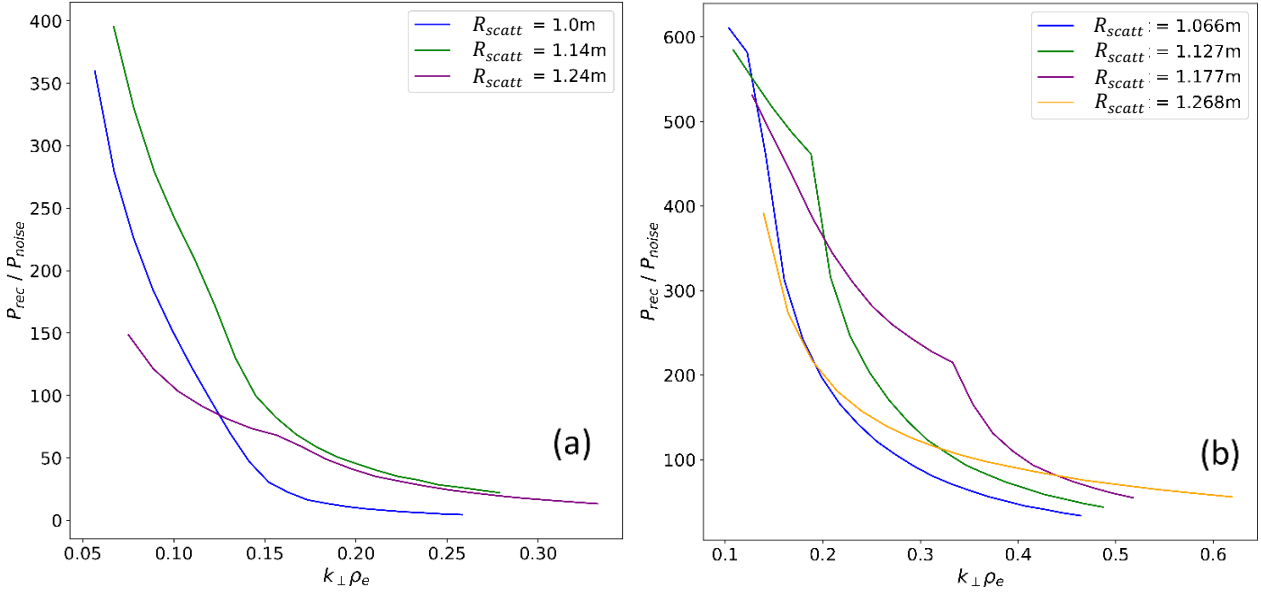

**Figure 12.** Received power to noise ratio for (a) the MAST-U sample high-beta equilibrium and (b) the MAST 022769 shot equilibrium using equation 1 and the localisation estimates $L_{FWHM}$ in figures 9b and 10b.

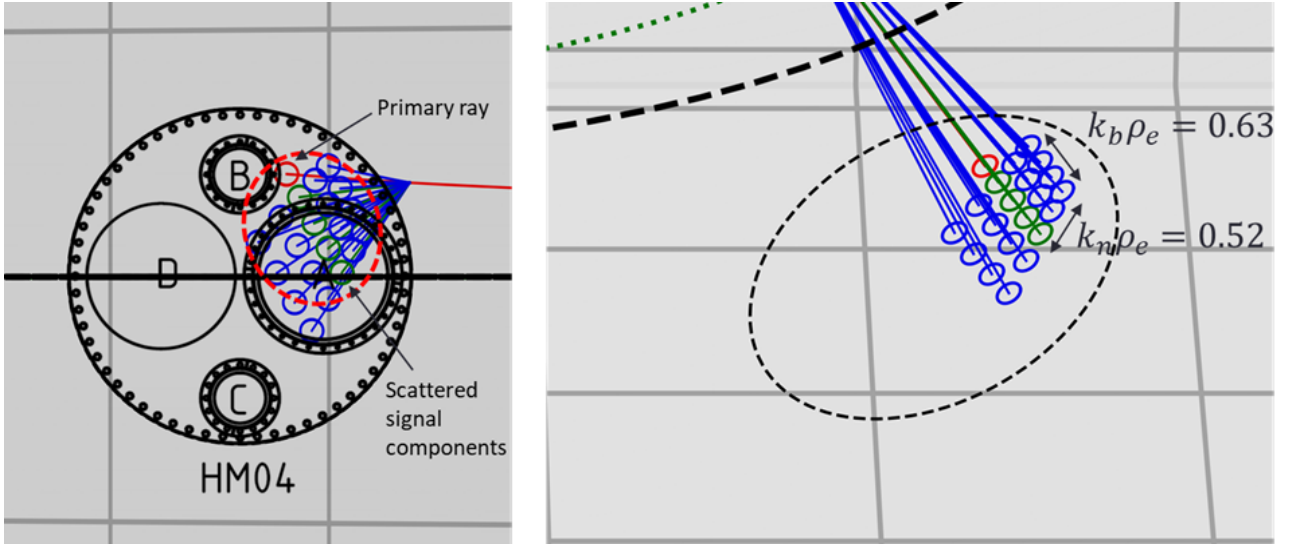

**Figure 13.** Projections of strictly bi-normal (green) and normal + bi-normal (blue) turbulence wavenumber contributions to scattered spectrum for $k_b = 3.95 \times 10^2, 8.52 \times 10^2, 1.31 \times 10^3, 1.77 \times 10^3$ m$^{-1}$ (green) and $k_n$ = -1.5 $\times 10^3$, -7.5 $\times 10^2$, 0.0, 7.5 $\times 10^2$, 1.5 $\times 10^3$ m$^{-1}$ (blue)

In both cases the signal to noise ratio drops to a minimum of around 10 for higher $k_\perp \rho_e$ values (larger scattering angles). In section 3.4, the scattered power due to a finite Gaussian beam incident on a simulated turbulence spectrum $\delta \hat{n}_e(\boldsymbol{k}, \omega)$ is calculated using equation 2, providing realistic estimates for the power spectrum received on each channel of the instrument detector array.

Although the MAST-U high-k scattering diagnostic is primarily optimised to measure bi-normal aligned electron scale turbulence, due to the pitch rotation capabilities of the receiving carriage and focussing optics, it is possible to rotate the detector channels outside strictly bi-normal alignment and measure a combination of bi-normal $k_b$ and normal $k_n$ turbulence wavevectors on each channel from a given radial scattering coordinate. Figure 13 illustrates the scattered channel distributions corresponding to a bi-normal wavevector range of $k_b \rho_e$ = 0 - 0.63 and a normal wavevector range of $\pm k_n \rho_e$ = 0.52. It is evident that some of the scattered components are outside the limits of the elliptical receiving window aperture, however for appropriate carriage rotation, a significant portion are measurable. This further extends the measurement capabilities of the instrument in characterising the radially dependent electron scale turbulence spectrum, as will be further illustrated in section 3.4.

| $r/a$ | 0.4 | 0.5 | 0.6 | 0.8 |
|---|---|---|---|---|
| $q$ | 1.04 | 1.1 | 1.2 | 2.5 |
| $\hat{s}$ | 0.06 | 0.34 | 1.1 | 4.63 |
| $\kappa$ | 1.41 | 1.41 | 1.42 | 1.49 |
| $\kappa'$ | -0.02 | 0.04 | 0.16 | 0.53 |
| $\delta$ | 0.10 | 0.16 | 0.14 | 0.23 |
| $\delta'$ | 0.29 | 0.22 | 0.24 | 0.90 |
| $\Delta'$ | -0.09 | -0.13 | -0.18 | -0.35 |
| $\beta_e$ | 0.07 | 0.06 | 0.05 | 0.03 |
| $\beta'$ | -0.53 | -0.57 | -0.49 | -0.29 |
| $a/L_{n_e}$ | 0.4 | 0.2 | 0.3 | 0.06 |
| $a/L_{T_e}$ | 1.6 | 2.1 | 2.2 | 3.1 |
| $a/L_{T_i}$ | 1.3 | 1.7 | 2.0 | 3.7 |

**Table 1.** Miller parametrisation of the radial surface at $r/a \in \{0.4, 0.5, 0.6, 0.8\}$ of the MAST equilibrium shot 22769. The parameters listed are the safety factor $q$, the magnetic shear $\hat{s} = \frac{r}{q}\frac{dq}{dr}$, the elongation $\kappa$ and its radial derivative $\kappa'$, the triangularity $\delta$ and its radial derivative $\delta'$, the Shafranov shift $\Delta'$, $\beta_e = 2\mu_0 p_e/B_0^2$ (where $p_e$ is the electron pressure). Also listed the local logarithmic radial gradient of electron density, electron temperature and ion temperature.

## 3.3 Electron scale gyrokinetic analysis of the reference MAST case

In this section, we describe electron scale gyrokinetic simulations used to provide realistic turbulence fluctuation maps used for synthetic diagnostic development in section 3.4. We briefly describe the local (flux-tube) gyrokinetic simulations performed at different radial locations of the MAST reference case (shot 022769). In particular, we consider four radial surfaces at $r/a = 0.4, r/a = 0.5, r/a = 0.6$, and $r/a = 0.8$. The choice of this MAST case is partially motivated by a previous gyrokinetic analysis of 022769 [40] that shows dominant transport from ETG turbulence at $r/a = 0.5$ and $r/a = 0.6$, which is comparable to the experimental heat flux value [41]. The gyrokinetic analysis [40] reveals the presence of an MTM (Microtearing Mode) ion scale instability at these surfaces in the binormal direction. This instability is neglected here however as the proposed diagnostic has been optimised primarily for electron scale measurements in the binormal direction. MTMs do however have a very fine radial structure (high $k_x$), which may be detectable with the extended normal $k_n$ measurement capabilities described in the previous section. The local linear and nonlinear gyrokinetic simulations considered in this work are performed using the CGYRO code [42]. The numerical resolution considered in linear simulations in the parallel, radial, velocity and pitch-angle direction is $(n_\theta, n_r, n_v, n_\xi) = (32, 32, 10, 24)$, where the pitch-angle is defined as $\xi = v_\parallel/v$, with $v_\parallel$ the velocity component parallel to the equilibrium magnetic field and $v$ the total velocity. Two species, electrons and deuterium, are considered. Simulations are fully electromagnetic, i.e. they evolve electrostatic potential fluctuations, $\delta\phi$, as well as both perpendicular and parallel magnetic fluctuations, $\delta A_\parallel$ and $\delta B_\parallel$. The value of local parameters obtained from a Miller parametrisation at each surface of the MAST 022769 shot [41], obtained using the Pyrokinetics Python library [43, 44] is reported in Table 1. Further details on the equilibrium and profiles are detailed in M. Valovic et al [41].

Figure 14 shows the growth rate and mode frequency as a function of the wavevector component $k_y\rho_s$, where $\rho_s = c_s/\Omega_D$ is the sound ion Larmor radius, with $c_s = \sqrt{T_e/m_D}$, $\Omega_D = eB_0/m_D$, $T_e$ the electron temperature on the chosen radial surface, $B_0$ is the total magnetic field at the centre of the chosen flux surface, and $m_D$ the deuterium mass. The growth rate and the mode frequency are normalized to $c_s/a$, evaluated on the corresponding radial surface. The maximum (normalized) growth rate value is achieved at $r/a = 0.8$, consistent with the higher electron temperature gradient present at this location. A marginally stable ETG instability is found at $r/a = 0.4$. The frequency is negative (phase velocity in the electron diamagnetic direction) and proportional to $k_y$, as expected from ETG instability. The parallel mode structure of $\delta\phi$ and $\delta A_\parallel$ at $k_y\rho_s \simeq 16$ of the $r/a = 0.5$ surface is shown in Figure 14. The amplitude of $e\delta\phi/T_e$ significantly exceeds that of $\delta A_\parallel/(\rho_s B_0)$, thus supporting the electrostatic nature of the underlying micro-instability.

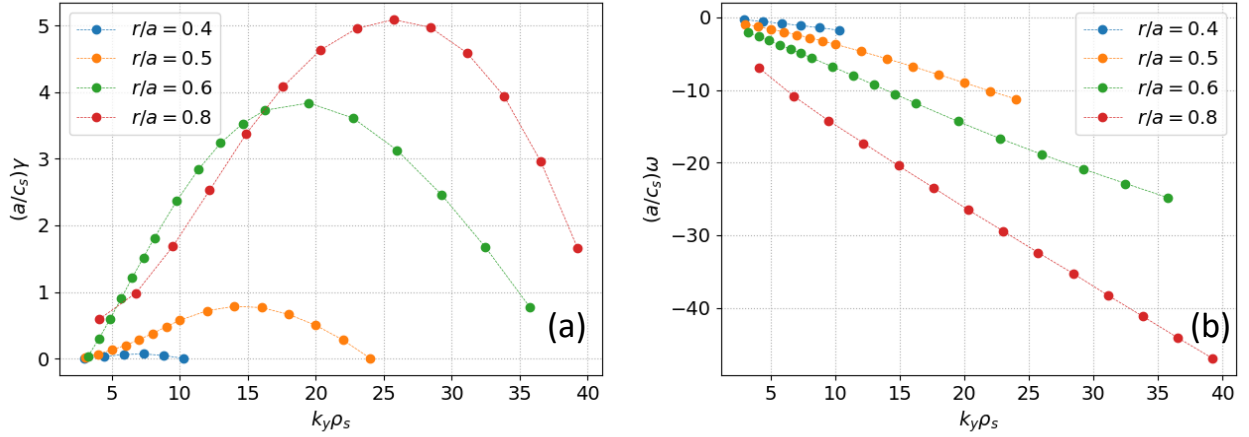

**Figure 14.** Growth rate (a) and mode frequency (b) of the electron scale instability as a function of the bi-normal wavevector $k_y\rho_s$ from linear gyrokinetic simulations at $r/a \in \{0.4, 0.5, 0.6, 0.8\}$. The growth rate and the mode frequency values are normalized to $c_s/a$ evaluated on the corresponding radial surface.

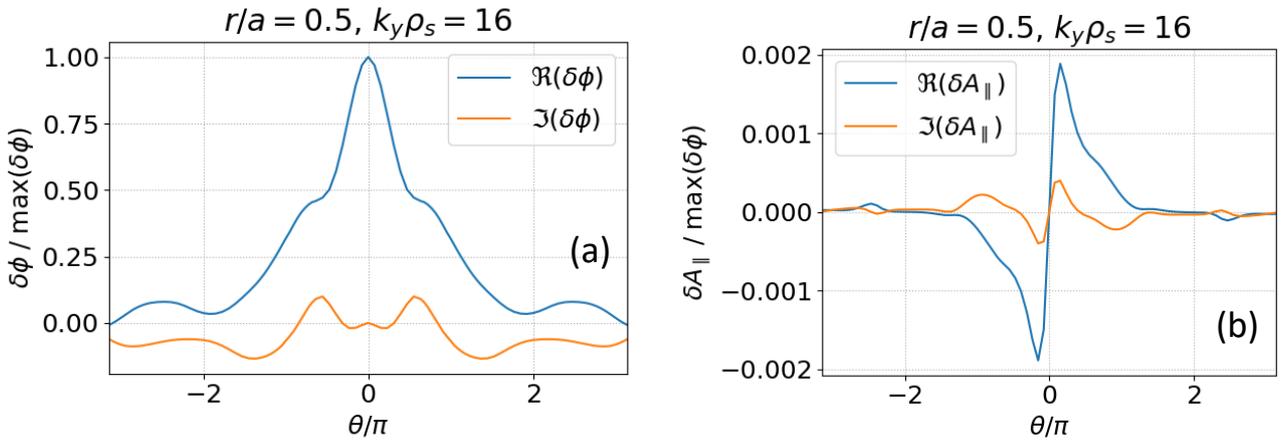

**Figure 15.** Parallel mode structure of $e\delta\phi/(\rho_* T_e)$ (a) and $\delta A_\parallel/(\rho_* \rho_s B_0)$ (b) corresponding to the $k_y\rho_s \simeq 26$ mode of the $r/a \simeq 0.8$ surface. Both $\delta\phi$ and $\delta A_\parallel$ are normalised to $\max \delta\phi$.

The nonlinear simulations were performed with the numerical resolution $(n_\theta, n_v, n_\xi) = (32, 10, 24)$ considering a $k_y\rho_s$ range that covers the linear electron scale instability spectrum in Figure 14 (see Table 2 for the numerical resolution used in $k_y$ and $k_x$). The equilibrium flow shear is not included in these nonlinear simulations. Flow shear was included in previous GK simulations of ETG turbulence in MAST and had little impact on the electron-scale turbulence [23].

The saturated heat flux value (normalised to the local gyro-Bohm heat flux $Q_{gB} = \rho_*^2 n_e T_e c_s$) obtained from the nonlinear simulations is shown in Figure 16 as a function of the radial position. As expected from the linear analysis, the heat flux driven by the ETG instability increases with radius. Figure 16 shows also the relative density fluctuation amplitude from ETG turbulence at the outboard midplane as a function of radius

$$\frac{\delta n_e}{n_e} = \frac{1}{n_e} \left\langle \sqrt{\sum_{k_x,k_y} |\delta n_e(k_x,k_y,\theta=0,t)|^2} \right\rangle_T$$

where $\langle \cdot \rangle_T$ denotes the time average performed over the last 30% of the total simulation time. We note that the density fluctuation amplitude increases with radius, similarly to the heat flux. Figure 16 shows also a snapshot of the electron density fluctuations at the outboard midplane and $r/a = 0.5$ taken at the last simulation time. The presence of radially elongated streamers is clearly visible in Figure 16, thus revealing a strong anisotropy of the turbulence between the radial and bi-normal directions, discussed in more detailed in the following sections. This anisotropy is observed at all the radial locations considered in this work and is consistent with previous observations [23].

| $r/a$ | $n_{k_y}$ | $n_{k_x}$ | $\Delta k_y\rho_s$ | $\Delta k_x\rho_s$ |
|---|---|---|---|---|
| 0.4 | 32 | 128 | 1.5 | 0.56 |
| 0.5 | 16 | 128 | 2.0 | 1.1 |
| 0.6 | 16 | 64 | 2.3 | 1.9 |
| 0.8 | 32 | 128 | 2.5 | 0.76 |

**Table 2.** Numerical resolution of the $(k_x, k_y)$ grid used in the nonlinear simulations at different radial positions.

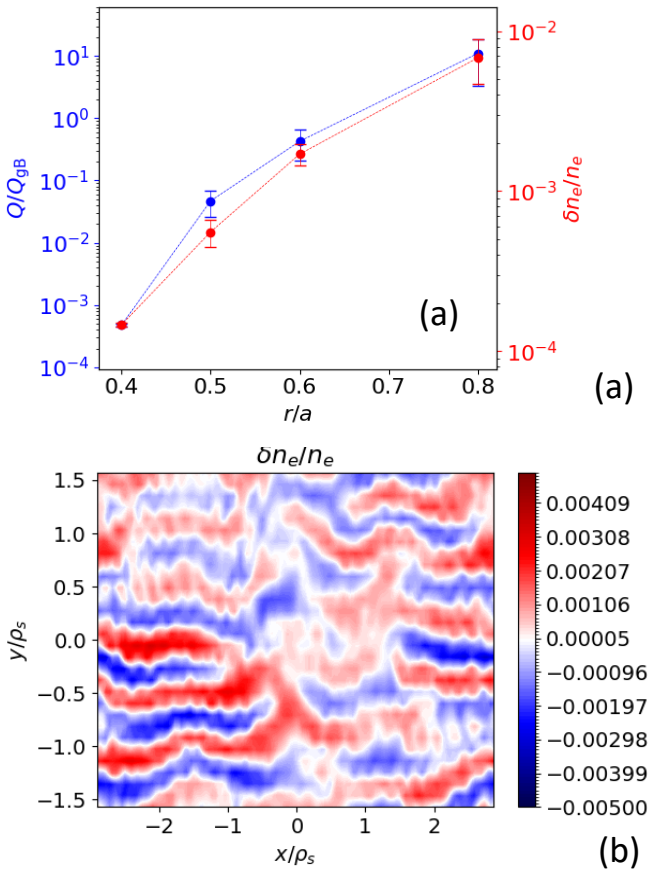

**Figure 16.** (a) Total heat flux normalised to $Q_{gB}$ (blue) and relative electron density fluctuation amplitude at the outboard midplane (red) as a function of radius. (b) Snapshot of the electron density fluctuations at the outboard midplane taken at the last simulation time of the simulation at $r/a = 0.5$.

*3.4 Synthetic high-k diagnostic for projecting high-k scattering measurements on MAST-U*

The predicted density fluctuation spectra from the CGYRO electron scale gyrokinetic simulations are used to predict the measured scattered power spectra from the proposed high-k scattering system on MAST-U. A synthetic / numerical implementation of the proposed high-k diagnostic is used for this purpose. The synthetic diagnostic employed here has recently been implemented into the Pyrokinetics framework [43, 44]. Consistent with the Pyrokinetics framework, the synthetic high-k diagnostic code is independent of the gyrokinetic code used to generate the density fluctuation spectra.

The high-k synthetic diagnostic employed here introduces the dependence of the scattered power on the density fluctuation power spectrum and on the spatial resolution $L_{FWHM}$. As we saw in section 2.1, the dependence on the spatial resolution is important, since the spatial resolution has explicit dependence on the measured $\boldsymbol{k}_\perp$ (see equation [3]). The dependence of the high-k measurement localization function is normally characterised by $L \propto 1/k_\perp$, which was absent in the previous synthetic high-k diagnostic analyses upon which this work is built [45, 46, 47]. This dependence is important as it can directly affect the physical interpretation of the measurement, as discussed below.

The synthetic high-k diagnostic takes as input the measured wavevector $\boldsymbol{k}_\perp$ and the spatial location of scattering. The second step is to run a nonlinear gyrokinetic simulation that adequately resolves the measured $\boldsymbol{k}_\perp$ at the given radial location of scattering. Therefore, it is a post-processing tool to the nonlinear gyrokinetic simulations that predicts the frequency and wavenumber spectrum given a set of selected scattered wavenumbers (here provided by Scotty [36]) and the simulated turbulence fluctuations. In what follows, we describe the implementation of the synthetic high-k diagnostic.

The starting point for developing the synthetic diagnostic is the Fourier expansion of the electron density fluctuation field $\delta n_e$ generated by a local gyrokinetic code. In local gyrokinetics, fluctuating fields such as $\delta n_e$ are represented as follows

$$\delta n_e(x,y,\theta,t) = \sum_{k_x,k_y} \delta \hat{n}_e(k_x,k_y,\theta,t)\exp(ik_x x + ik_y y) \quad (7)$$

where $k_x$ and $k_y$ are the wavenumber components of $\boldsymbol{k}_\perp$ as defined in Pyrokinetics, normalized by the reference magnetic field $B_0$, and $x$ and $y$ are the respective conjugate spatial directions perpendicular to the background magnetic field vector $\boldsymbol{B}$. It is important to note that the internally defined $k_x$ and $k_y$ wavenumber components in gyrokinetic codes do not generally correspond to the normal and bi-normal components $k_n$ and $k_b$. One needs a mapping between the two wavenumber definitions. In the synthetic high-k diagnostic, we map the selected $k_n$ and $k_b$ components obtained from Scotty to the internal $k_x$ and $k_y$ components used in Pyrokinetics, as shown in recent works [47, 48]. This allows one to identify which wavenumber components $k_x$ and $k_y$ from a gyrokinetic code correspond to a specific diagnostic measurement configuration.

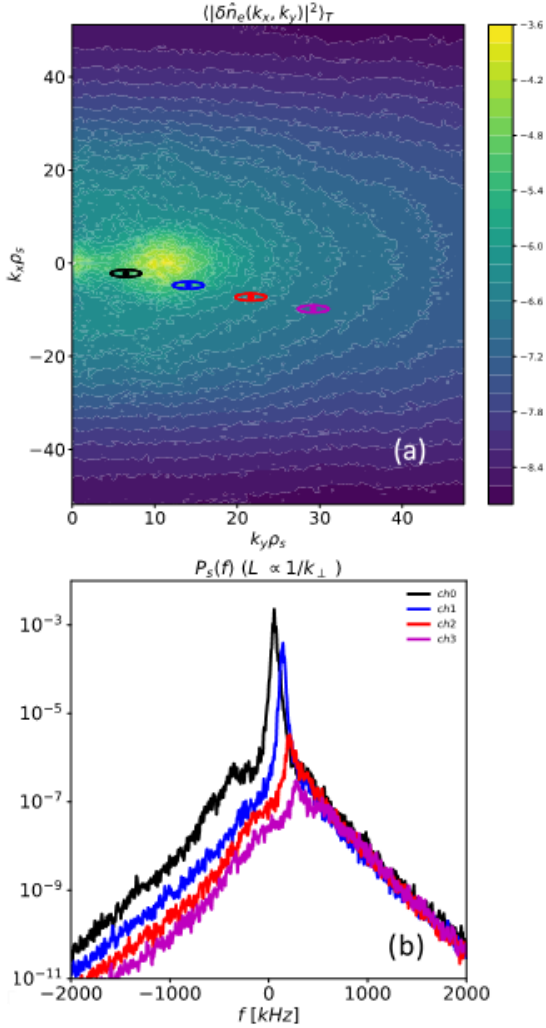

**Figure 17.** (a) Density fluctuation wavenumber power spectrum from the CGYRO simulation at *r/a* = 0.5, for a scattering configuration scanning the $k_b$ component of the turbulence spectrum. Coloured dots and ellipses represent the selected $k_\perp$ and its resolution for the different channels on the scattering diagnostic, respectively. (b) Frequency spectrum of the turbulent electron density fluctuation power corresponding to the selected wavenumbers in (a).

In addition to mapping the specific ($k_n$, $k_b$) pair to ($k_x$, $k_y$), one needs to map the diagnostic wavenumber resolution $\Delta k_\perp$ to $\Delta k_x$ and $\Delta k_y$. The reader is referred to previous work [46] for additional details on the wavenumber mapping between ($k_n$, $k_b$) and ($k_x$, $k_y$), as well as the corresponding wavenumber resolution. We note that, while the wavenumber components $k_x$ and $k_y$ are code dependent and not general, the $k_n$ and $k_b$ components remain universal, further justifying their use within the Pyrokinetics standardized framework for gyrokinetic simulations.

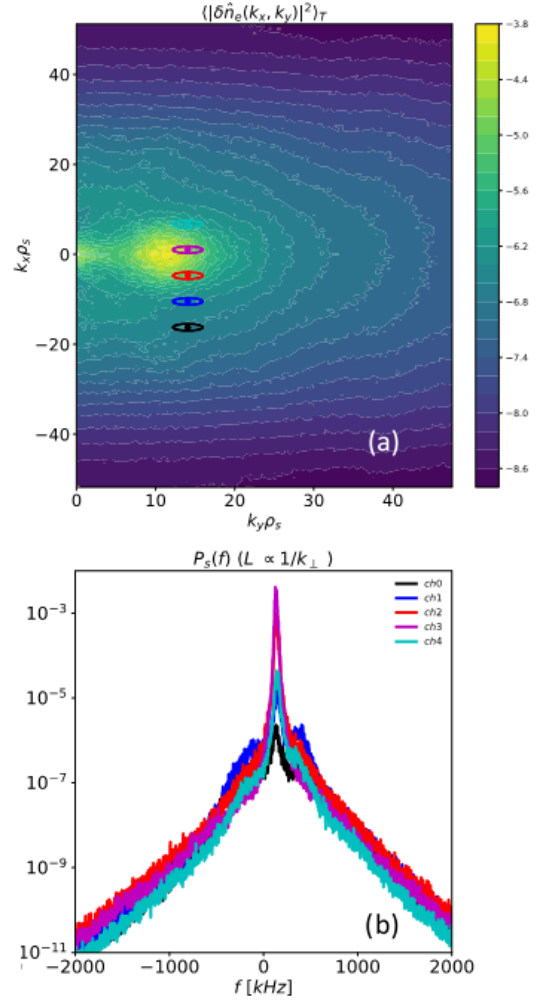

**Figure 18.** Similar to figure 17, this time performing a scan in the $k_n$ component of the turbulence spectrum. (a) Density fluctuation wavenumber power spectrum from the CGYRO simulation at *r/a* = 0.5. (b) Frequency spectrum of the turbulent electron density fluctuation power corresponding to the selected wavenumbers in (a).

Simulated electron-scale turbulence from CGYRO is used in conjunction with a synthetic high-k diagnostic to enable quantitative projections of future high-k turbulence measurements in MAST-U. A specific example of four mapped wavenumber pairs ($k_n$, $k_b$) to ($k_x$, $k_y$) is given by the coloured dots on Figure 17a. These correspond to the four channels of the high-k scattering diagnostic for the radial location of *r/a* = 0.5, and are overlayed to the 2D density fluctuation power spectrum $\langle|\delta \hat{n}_e|^2(k_x, k_y)\rangle_T$ that has been numerically computed by CGYRO. The four channels are set in a configuration to scan the bi-normal $k_b$ component of the turbulence, while maintaining a fixed $k_n = 0$. We refer to this configuration as a $k_b$-scan. Although $k_n = 0$ for the four channels, the flux surface geometry and out of midplane scattering location introduces a different $k_x$ for each channel (see Ruiz Ruiz PPCF 2020/2022 for additional details [45, 47]). The wavenumber resolutions $\Delta k_x$ and $\Delta k_y$ are represented by the ellipses surrounding each coloured point in Figure 17a, and correspond to the $1/e^2$ power contributions to the scattered power.

Figure 17a shows that the currently proposed configuration of the high-k scattering instrument can probe the density fluctuation spectrum close to the peak of spectrum by performing a scan in the bi-normal component $k_b$ of the turbulence. This spectral peak is generally attributed to radially elongated and poloidally thin turbulent structures known as streamers [20, 21, 22]. Figure 17b shows the frequency spectral power (in arbitrary units) of the turbulent density fluctuations corresponding to each coloured point in Figure 17a. As expected, the higher wavenumber channels exhibit a decreasing fluctuation power. Additionally, note how the higher wavenumbers also exhibit a higher frequency $f$. This is due to the nature of the turbulent fluctuations, which exhibit a higher frequency for higher wavenumbers. This means that turbulent structures of smaller physical dimensions propagate faster than those of larger physical dimensions. The propagation of the turbulence fluctuations in the plasma frame propagation is also commonly denominated as the phase velocity, given by $f/k$. We note here that the frequency shift from Figure 17b is not a Doppler shift, as the Doppler shift is not included in the current analysis. Adding a Doppler shift to Figure 17b would increase the frequency response required of the diagnostic, but would have no impact on the measured $\boldsymbol{k}_\perp$ nor on the total scattered power. Figure 17b could be quantitatively compared to experimental measurements from the high-k scattering instrument.

Figure 18 shows the predicted measurement range of the proposed diagnostic by performing a scan in the normal component $k_n$ of the turbulence. We refer to this configuration as a $k_n$-scan. This measurement configuration is designed to select a finite $k_b = 8.78$ cm$^{-1}$ that is close to the driving, or injection scale of the turbulence, which is also where ETG streamers are predicted to exist. The scattering configuration to perform a $k_n$-scan requires consideration of the physical diagnostic configuration. In the proposed high-k scattering instrument, a $k_n$ scan is made possible by rotating the frame of detectors from strictly bi-normal alignment with a finite offset angle. Each channel of the linear detector array will then align with a unique combination of $k_b$ and $k_n$ values for a given scattering radius. As can be seen from Figure 18, the scattering configuration of the diagnostic required to perform a $k_n$-scan predicts the measurement of the $k_x$ dependence of the electron density fluctuation spectrum around the spectral peak due to ETG streamers. Taken together, Figures 17 and 18 indicate that the MAST-U high-k scattering diagnostic will be successful at measuring both the $k_x$ and the $k_y$ dependence of the density fluctuation spectrum around the peak wavenumber of the radial streamers.

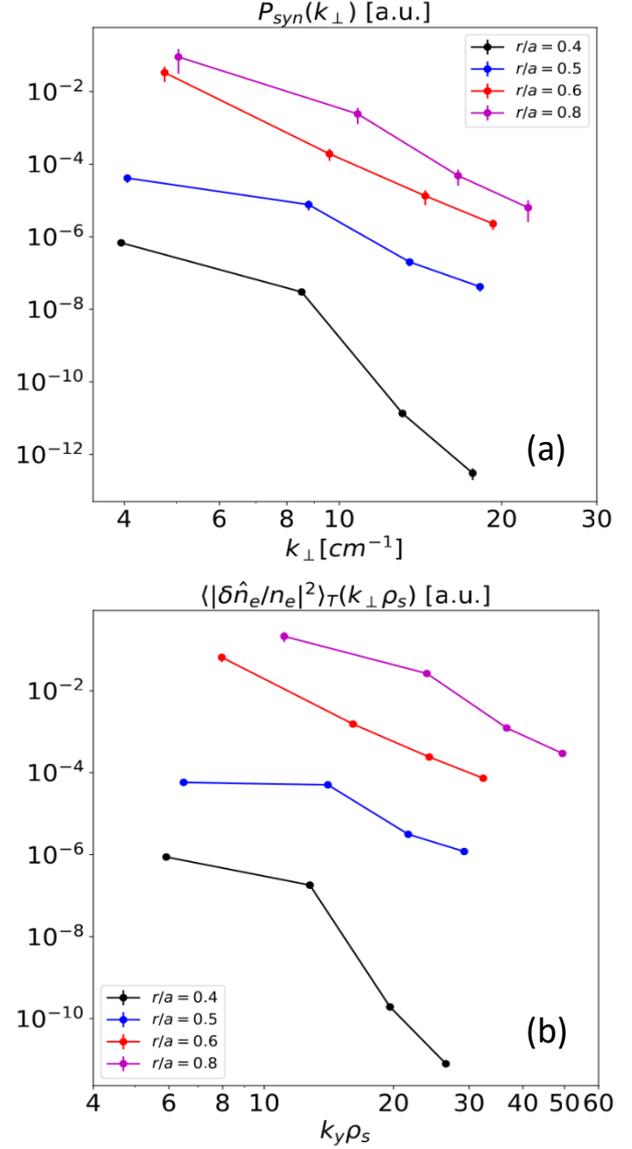

**Figure 19.** (a) Synthetic, predicted scattered power by high-k synthetic diagnostic and (b) electron density fluctuation power spectrum. A scan in the bi-normal $k_b$ component of the turbulence was performed for $r/a = 0.4, 0.5, 0.6$ and $0.8$. The large variation of the scattered power is due to the variation of the turbulence intensity with minor radius (see figure 16).

Despite the apparent limitation to four channels, the measurement could be populated with additional points corresponding to different k-values measured over repeated plasma discharges via collective linear translation of the Schottky diode detector array and discrete receiving antennae.

The frequency response of the turbulence for the $k_n$-scan configuration is given in Figure 18b. Note how the frequency dependence of the turbulent spectrum exhibits a similar dependence for the different selected $k_n$. Contrary to Figure 17b, where a clear shift of the frequency spectrum while scanning $k_b$ is observable, Figure 18b does not show a shift towards larger frequencies when scanning $k_n$. This is because the dominant contribution to the phase velocity of the turbulence is in the bi-normal direction, along $k_b$. Turbulent structures tend to propagate within the flux surface in the bi-normal direction much faster than in the radial direction, as confirmed by nonlinear gyrokinetic simulations [42, 49]. Figure 18b shows the expected clear dependence of the total spectral amplitude in the frequency power spectrum. This dependence proved beneficial in the development and interpretation of Figure 19.

A measurement of the total predicted scattered power and the corresponding electron density fluctuation power can also be performed by using the diagnostic. Figure 19a shows the total, synthetic scattered power $P_{syn}$ as a function of the selected $k_\perp$ from the proposed diagnostic for the $k_b$-scan configuration. The scattered power varies by several orders of magnitude with minor radius, as well as with the selected $k_\perp$. This large variation is primarily due to the variation of the turbulence fluctuation intensity with minor radius, which depends strongly on the gradients of the background plasma profiles, especially on the electron temperature gradient as is visible in Figure 14. Figure 19b shows the predicted electron density fluctuation power spectrum $\langle |\delta \hat{n}_e|^2(k_\perp)/n_e^2\rangle_T$ as a function of $k_y\rho_s$ for the $k_b$-scan. Following equation (2), the measured electron density fluctuation spectrum can be inferred from the synthetic scattered power $P_{syn}$ by $\langle |\delta \hat{n}_e|^2(k_\perp)/n_e^2\rangle_T \propto k_\perp^2 P_{syn}$, where the additional factor of $k_\perp^2$ has its origin in the $k_\perp$ dependence of the diagnostic spatial resolution $L$, see equation (2). Note how the peak $\langle |\delta \hat{n}_e|^2(k_\perp)/n_e^2\rangle_T$ around $k_y\rho_s \approx 10$ can be inferred for the radial location $r/a = 0.5$ (blue curve in Figure 19b, corresponding to Figures 17 and 18). For larger $k_y\rho_s$, we observe a decrease in the spectral power with wavenumber, characteristic of the inertial range of the turbulent cascade. This behaviour is not observed at all radial locations, and would need to be analysed in detail on a case-by-case basis.

In addition to calculating the peak wavenumber $k_b$ (or $k_y\rho_s$) corresponding to ETG streamers, the aspect ratio $(k_y/k_x)$ of the turbulent eddies could also be calculated from such a measurement. Having measured $\langle |\delta \hat{n}_e|^2(k_\perp)/n_e^2\rangle_T$, one could calculate the $1/e^2$ spectral widths $w_{k_x}$ and $w_{k_y}$ that characterise the spectral decay of the turbulent spectrum, respectively in the $k_x$ and in the $k_y$ directions. The aspect ratio of the turbulent eddies in the perpendicular direction to the background magnetic field can be calculated by the ratio of the radial correlation length to the bi-normal correlation lengths, which can in turn be related to the spectral widths as follows $l_r/l_b \approx w_{k_y}/w_{k_x}$ [46, 47]. This analysis is not the object of this publication, and its implementation is left to the analysis of data from a future measurement using the proposed high-k scattering instrument. Figure 19 suggests that the diagnostic will be able to quantitatively distinguish from conditions of strong and weak ETG turbulence drive in MAST-U, as well as to determine intrinsic characteristics of the turbulence, which is one of the main objectives of the diagnostic.

## 4. Discussion

A high-k electron scale turbulence scattering diagnostic model has been designed for future implementation on MAST-U. The instrument operates in a collective scattering regime $(1/k_\perp \lambda_{De} \geq 1)$ based on the principles of Bragg scattering, and is primarily designed to diagnose the ETG turbulence wavenumber spectrum in the bi-normal direction with adjustable spatial localisation from the plasma core to the edge pedestal region. Due to an operating frequency of 376 GHz, measurement within the core plasma is possible under all operational conditions of MAST-U. A highly flexible rotatable and translatable receiving optics carriage containing four scattering detector channels enables precise alignment with the bi-normal direction to be maintained whilst adjusting the radial location (and imaging focus) of the receiving optics. Rotational adjustment also facilitates measurement of a range of bi-normal $k_b$ and normal $k_n$ turbulence wavenumber combinations through rotational misalignment from strictly bi-normal incidence. For a sample high-beta MAST-U equilibrium, beam tracing simulations project the maximum normalised bi-normal wavenumber of measurement to be $k_\perp\rho_e \sim 0.34$ in the core and $k_\perp\rho_e \sim 0.43$ near the pedestal. For the reconstructed MAST shot 022769 with lower magnetic field, the maximum normalised bi-normal wavenumber of measurement is $k_\perp\rho_e \sim 0.6$ in the core and $k_\perp\rho_e \sim 0.79$ near the pedestal (these projections are aperture limited by the 250 mm × 290 mm elliptical receiving window). The instrument selectivity function along the path of scattering has been analysed using the formalism developed in section 2.1. This analysis combines the finite overlap of the incident and scattered Gaussian beams [38] with the rotational misalignment effect due to magnetic field pitch angle variation with radius [31, 37]. For the high-beta MAST-U equilibrium, the associated localisation length $L_{FWHM}$ varies between a maximum of ~0.33 m for $k_\perp\rho_e \sim 0.05$ in the core and a minimum of ~0.05 m for $k_\perp\rho_e > 0.25$ at all radial coordinates. For the 022769 MAST equilibrium, $L_{FWHM}$ ranges from ~0.4 m in the core for $k_\perp\rho_e \sim 0.1$ to ~0.08m for $k_\perp\rho_e > 0.45$. The localisation lengths have been used to estimate the scattered power assuming a single coherent density fluctuation with $dn_e/n_e \sim 4 \times 10^{-6}$. This yielded a maximum received power to detector noise ratio $P_{rec}/P_{noise}$ of ~400 → 600 for both sample equilibria at smallest $k_\perp\rho_e$ and a $P_{rec}/P_{noise}$ of ~10 at maximum $k_\perp\rho_e$. The minimum scattered power is therefore comfortably above a reasonable detection threshold.

In order to compare the measurement specifications of the diagnostic with a sample ETG dominated turbulence map at corresponding radial scattering coordinates, electron scale localised (flux-tube) gyrokinetic simulations were conducted using the CGYRO code for the 022769 MAST equilibrium. The simulations are fully electromagnetic and use two species, electrons and deuterium ions. Calculations of the ETG turbulence linear growth rate [40] show a maximum (normalized) growth rate value at $r/a = 0.8$, consistent with the higher electron temperature gradient at this radius. The correspondingly largest normalised turbulence wavenumber for peak growth is $k_y\rho_s = 26$ also occurring at $r/a = 0.8$. Only a marginal ETG instability is found at $r/a = 0.4$ with correspondingly lowest growth rate and lowest turbulence wavenumber for peak growth at $k_y\rho_s = 7$. The nonlinear gyrokinetic simulations were conducted for $r/a = 0.4, 0.5, 0.6$ and $0.8$ using a $k_y\rho_s$ range determined by the electron scale instability spectrum from the linear calculations. As expected, the heat flux driven by the ETG instability increases with radius as does the relative amplitude of the density fluctuations, with $dn_e/n_e$ ranging between $\sim 1 \times 10^{-4}$ in the core to $\sim 1 \times 10^{-2}$ at the pedestal. This is at least $25 \times$ greater than the $\delta n_e/n_e = 4 \times 10^{-6}$ used in the scattered power calculations, equating to a received power that is $625 \times$ greater than previous minimum estimates.

In order to properly scale and project the measured normalised turbulence wavenumber specifications of the high-k scattering diagnostic to CGYRO ETG turbulence maps, a synthetic diagnostic framework was used [45]. This framework has been implemented within the Pyrokinetics Python library [43, 44] and enables calibration of the measured wavenumber specifications into gyrokinetic field-aligned coordinates, accounting for the plasma elongation and Shafranov shift [45]. The synthetic diagnostic framework also enables calculation of the scattered power frequency spectrum for each channel of the scattering diagnostic, using the wavenumber specifications and localisation lengths derived from ray-tracing to enable quantitative projections of future high-k turbulence measurements on MAST-U. Wavenumber coordinate mapping between the scattering instrument projections and CGYRO for each of the radial coordinates shows that the lowest 2 channels in $k_\perp\rho_e$ are coincident with the peaks in the turbulence spectra in all cases. This indicates that the $k_\perp\rho_e$ range measurable by the proposed high-k diagnostic would be more than adequate to resolve the simulated turbulence spectra, whilst providing additional channels for extended wavenumber coverage and relative measurements at lower expected powers. The additional channels can also facilitate off-bi-normal rotational misalignment for measurements combining bi-normal and normal turbulence wavenumber contributions. The composite synthetic power spectra illustrate that variations in the scattered power received per channel and for different $r/a$ values are clearly discernible, allowing mapping of the electron scale turbulence spectra both by wavenumber and spatial coordinate. The diagnostic will therefore be able to quantitatively distinguish between conditions of strong and weak ETG turbulence on MAST-U, whilst providing intrinsic characteristics of the turbulence at spatial coordinates from the core to the pedestal.

## 5. Conclusions

We have used a synthetic diagnostic approach to design a highly optimised yet flexible electron-scale turbulence diagnostic, using powerful modelling tools to predict the sensitivity and spectral range of measurement whilst diagnosing both normal and binormal turbulence wavenumbers. The morphology of the spherical tokamak plasma (notable, the magnetic field pitch rotation with radius) enhances the localization of measurement and provides further motivation for applying this instrument to MAST-U plasmas. There is however no reason why the same numerical optimisation and analysis technique could not be applied to a conventional tokamak. The proposed diagnostic opens up opportunities to study new regimes of turbulence and confinement, particularly in association with upcoming non-inductive, microwave based current drive experiments on MAST-U. These experiments are critical to the development of the future STEP (Spherical Tokamak for Energy Production) reactor and operational parameters. On MAST-U, the diagnostic expands on the capabilities of existing DBS diagnostics both in terms of turbulence wavenumber measurement range and core plasma access, while complementing ion-scale measurements of the BES diagnostic. It is also of particular relevance to the future STEP reactor due to its high operational frequency and resilient component design, making it highly suited to operation during burning plasma experiments where high-power microwaves are used for heating and current drive. Target physics problems include the diagnosis of ETG turbulence anisotropy / streamer formation, identifying cross-scale turbulence effects such as the suppression of electron-scale turbulence by ion-scale eddies [9] and the study of ETG turbulence in the presence of large scale (MHD) fluctuations (as expected in high-beta spherical tokamak plasmas) and Alfvénic instabilities which can themselves be destabilised by highly energetic particles (such as alpha particles) during future burning plasma experiments on STEP. The initial measurements of these effects on MAST-U will help benchmark the predictions of numerical models, and serve as a baseline for future comparative measurements of STEP turbulence [50, 51].

## 6. Acknowledgements

The authors are grateful for productive discussions with T. L. Rhodes at UCLA in the development of this work. The authors would also like to thank J. Harrison, R. Scannell and K. Hawkins at CCFE for their valuable input. The research was made possible by funding received from the EPSRC (Grant No. EP/R034737/1) and from A*STAR, via Green Seed Fund C231718014 and a SERC Central Research Fund. This work has been partially funded by the EPSRC Energy Programme (Grant No. EP/W006839/1). J. Ruiz-Ruiz was partially funded by EPSRC grant EP/W026341/1

## 7. Appendix

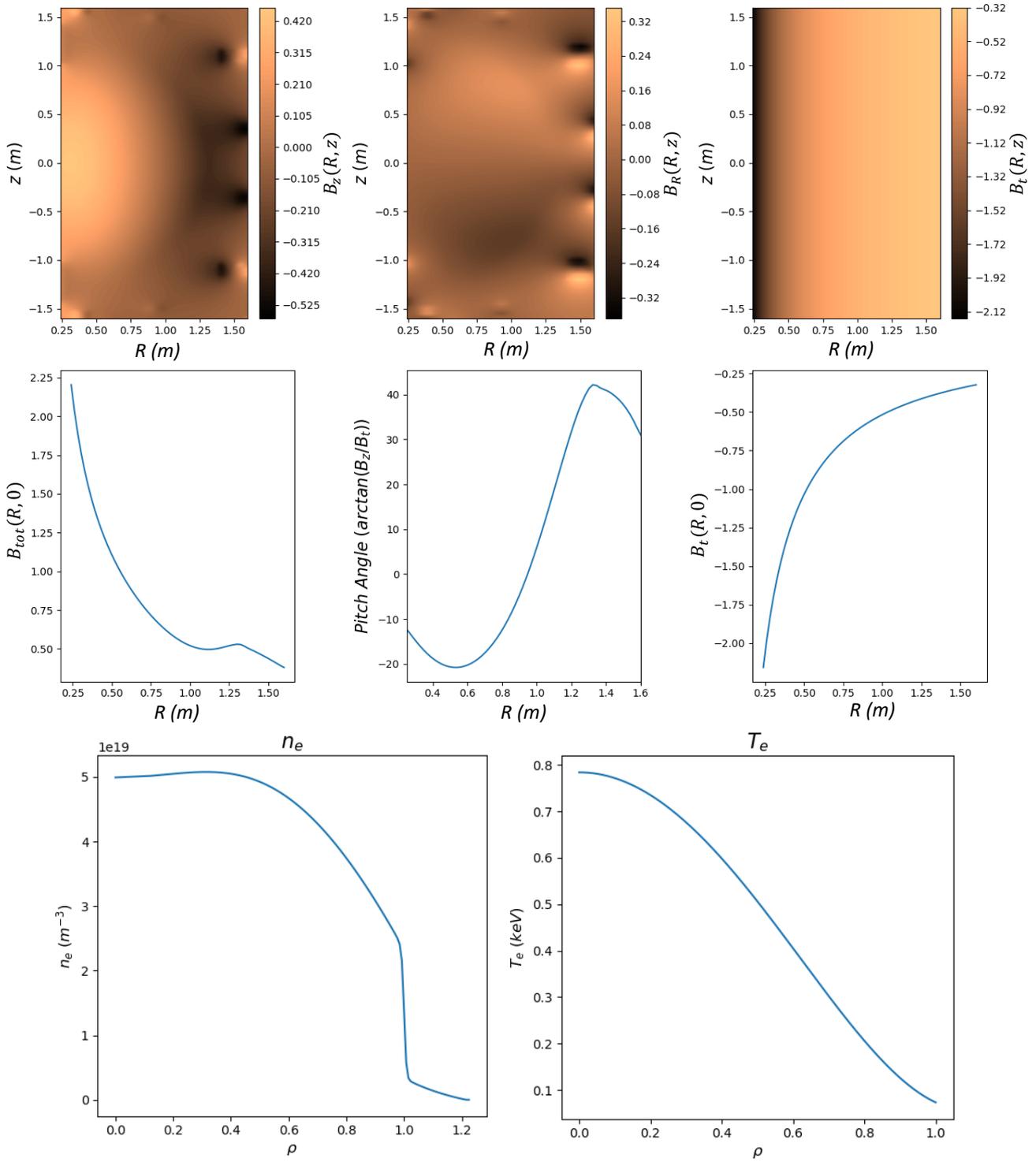

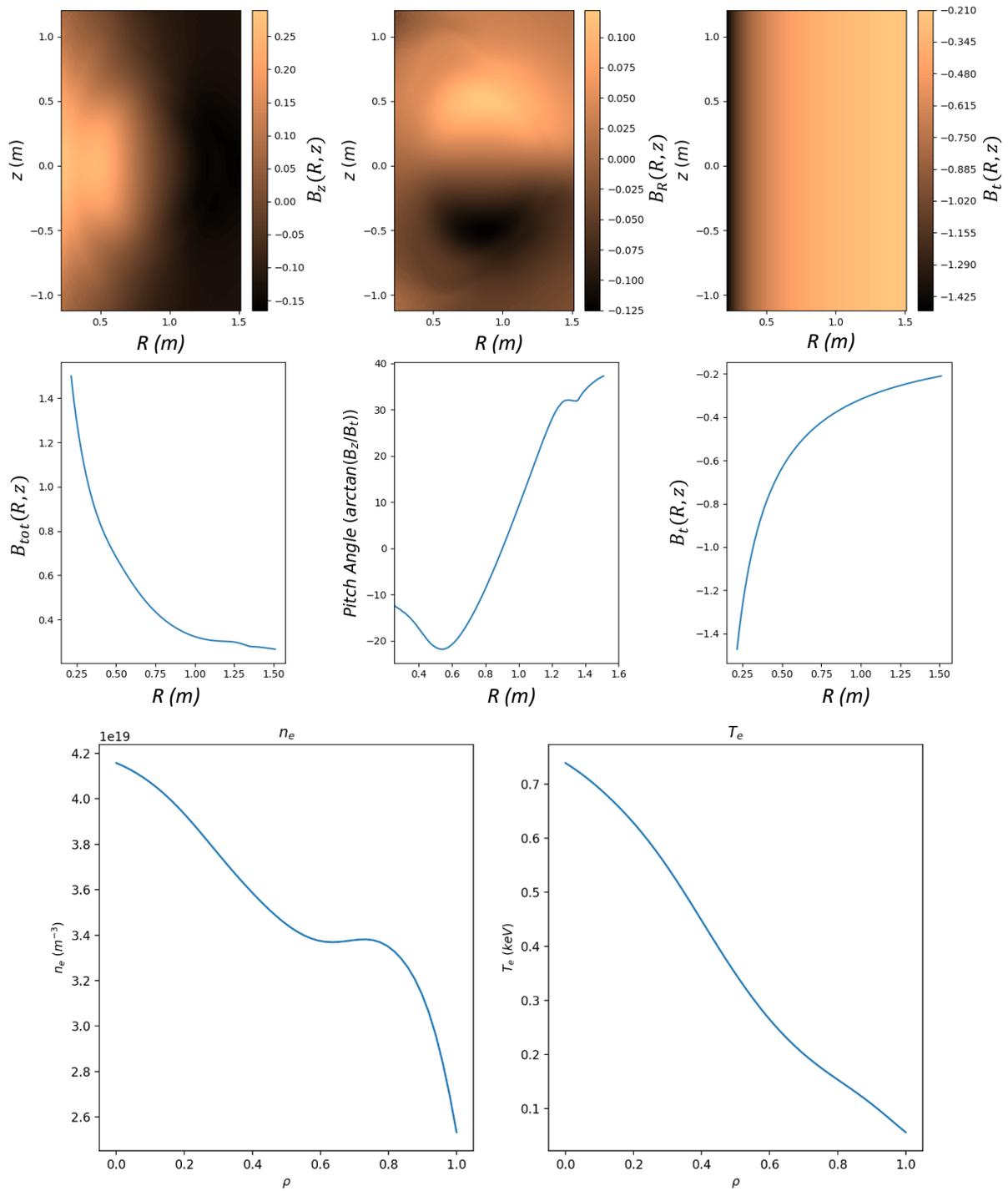